\documentclass[%
 reprint,
 amsmath,amssymb,
 prl,
]{revtex4-2}

\usepackage[english]{babel}

\usepackage{graphicx}
\usepackage{dcolumn}
\usepackage{bm}
\usepackage{hyperref}

\newcommand{\rvec}{ \mathbf{r} }

\newcommand{\nhat}{ \mathbf{\hat{n}} }

\begin{document}

\preprint{APS/123-QED}

\title{Phase behaviour of active particles in block copolymer melts}%

\author{Javier Diaz}
\affiliation{%
 CECAM, Centre Europ\'een de Calcul Atomique et Mol\'eculaire,
EPFL, \'Ecole Polytechnique F\'ed\'erale de Lausanne, \\
Batochime - Avenue Forel 2, 1015 Lausanne, Switzerland 
}
\author{Ignacio Pagonabarraga}%
 \email{ipagonabarraga@ub.edu}
\affiliation{%
 CECAM, Centre Europ\'een de Calcul Atomique et Mol\'eculaire,
EPFL, \'Ecole Polytechnique F\'ed\'erale de Lausanne, \\
Batochime - Avenue Forel 2, 1015 Lausanne, Switzerland 
}%
\affiliation{
Departament de F\'isica de la Mat\`eria Condensada, Universitat de Barcelona, Mart\'i i Franqu\'es 1, 08028 Barcelona, Spain 
}%
\affiliation{  
Universitat de Barcelona Institute of Complex Systems (UBICS), Universitat de Barcelona, 08028 Barcelona, Spain
}%

\date{\today}

\begin{abstract}
Block copolymer melts offer a unique template to control the position and alignment of nanoparticles due to their ability to self-assemble into periodic ordered structures. 
Active Brownian particles are shown to co-assemble with block copolymers leading to emergent organised structures. 
The block copolymer acts as a soft confinement that can control the self-propulsion of the active Brownian particles, both for interface-segregated and selective nanoparticles. 
At moderate activity regimes, active Brownian particles can form organised structures such as polarised trains or rotating vortexes. 
At high activity, the contrast in the polymeric and colloidal time scales can lead to particle swarms with distorted block copolymer morphology, due to the competition between the polymeric self-assembly and the active Brownian self-propulsion. 
\end{abstract}

\maketitle

Block copolymer (BCP) melts are ideal matrices to template  the location of passive nanoparticles (NPs), due to their ability to self-assembly into ordered mesophases. 
\cite{bates_block_1990,matsen_unifying_1996,thompson_predicting_2001,kim_creating_2007} such as lamellae of hexagonally ordered circles in two dimensions (2D).
This is due to the heterogeneity of the BCP chain, where monomers of chemically different types are grouped together into blocks.
Passive particles have been shown to form ordered structures when miscible with one of the BCP phases\cite{ploshnik_co-assembly_2010,ploshnik_hierarchical_2010,ploshnik_hierarchical_2013}, or decorating the interface\cite{kim_creating_2007}.
Furthermore, NPs can modify the BCP morphology by changing the effective composition of the BCP melt, leading to a rich phase behaviour of the hybrid co-assembled system\cite{huh_thermodynamic_2000,halevi_co-assembly_2014}. 
This work addresses both the structures that active particles can form mediated by the BCP template, as well changes in the BCP mesophase due to activity. 

Active particles are intrinsically out of equilibrium, as they consume energy from the surrounding medium to produce work. 
This work can be used, for instance, to allow particles to self-propel with a given direction. 
Suspensions of self-propelled particles have been shown to form a rich phase behaviour\cite{tailleur_statistical_2008,cates_motility-induced_2015,telezki_simulations_2020}, which has been largely modelled within the active Brownian particle (ABP) model, where spherical particles self-propel with a rotational diffusion time and excluded volume interactions\cite{hagen_brownian_2011,
romanczuk_active_2012,
cates_when_2013,
fily_athermal_2012,
redner_structure_2013,
bialke_microscopic_2013,
digregorio_full_2018}.  

Furthermore, the dispersion of active particles within complex medium has resulted in emergent organised structures\cite{bechinger_active_2016,Frangipane_invariance_2019}, such as the accumulation of ABPs at hard confining walls or the formation of vortexes of ABPs under circular confinement\cite{bricard_emergent_2015,lushi_fluid_2014,hernandez-ortiz_transport_2005,yang_aggregation_2014,wioland_confinement_2013}. 
While most of the literature covers the hard confinement of ABPs near solid and immobile interfaces, the use of soft walls permits the activity to deform the confining medium, such as the case of active filaments within vesicles \cite{kruse_asters_2004,peterson_vesicle_2021}. 
The geometry of obstacles or confinement has been shown to play a role in the accumulation of active particles in curved walls
\cite{angelani_geometrically_2010,mallory_curvature-induced_2014,wensink_controlling_2014}. 
Furthermore, inhomogeneous media can lead to localisation of active particles in the absence of hard confinement\cite{fernandez-rodriguez_feedback-controlled_2020}. 
These works motivate the use of BCP melts as templates to control the dispersion of active particles
that can result in emergent co-assembled BCP/ABPs morphologies. 

Active Janus particles at air/water interfaces have been shown to enhance their active persistence length by reducing their rotational diffusivity\cite{wang_enhanced_2015}, driven by the wetting of the particle\cite{wang_wetting_2016}. 
While  equilibrated flat air/water interfaces have been used to study the constrained collective behaviour of ABPs, in this work we exploit the intrinsic structure of the BCP interface.
NPs with chemically inhomogeneous surfaces have been successfully dispersed within BCP melts, leading to their achoring at interfaces\cite{yang_design_2017,yang_janus_2017}. 

In this work we study how the co-assembly of ABPs in BCP melts differs from passive NPs and the effect of activity in the emergence of structures for hybrid NP/BCP composites. 
We use a minimal model of ABPs coupled with a Cahn-Hilliard description of a BCP melt. 
The ability of BCP melts to self-assemble can be used as a case study of ABPs dispersion within mixtures with intrinsic ordering in the nanoscale.

In order to capture the overall phase behaviour of active NPs in BCP melts, we use a mesoscopic hybrid model that allows to reach relatively large length scales. 
The total free energy of the system is decomposed as $F_{tot}=F_{pol}+F_{cpl}+F_{cc}$, where $F_{pol}$ is the standard Ohta-Kawasaki\cite{ohta_equilibrium_1986} continuous  description of a BCP melt via the differences in concentration of A and B monomers $\psi=\phi_A-\phi_B+(1-2f_0)$.
The colloid-colloid contribution is a pairwise additive potential that prevents overlapping between ABPs. 
The colloid and the polymeric description are coupled with the interaction term $F_{cpl}=\sum_i \sigma \int d\textbf{r} \psi_c(r)\left[ \psi-\psi_0 \right]^2$ which introduces a coupling energetic scale $\sigma$ and a chemical affinity parameter $\psi_0$ that specifies the wetting of the ABP with the BCP. 
The shape and size of the particle is controlled by the tagged function\cite{tanaka_simulation_2000} $\psi_c$.

The dynamics of a system of $N_p$ ABPs with a radius $R$ is described  \textit{via} the overdamped Langevin equations
\begin{subequations}
\begin{equation}
\frac{d \textbf{r}_i}{dt}=
v_a\mathbf{\hat{n}}_i  +
\gamma_{t}^{-1} \left( \textbf{f}_i^{cc}+\textbf{f}_i^{cpl} \right) + \sqrt{2D_{t}}\mathbf{\xi}_{t}
\end{equation}
\begin{equation}
\frac{d \phi_i}{dt}=
\gamma_{r}^{-1} \left( M_i^{cc}+M_i^{cpl} \right) + \sqrt{2D_{r}}\xi_{r}
\end{equation}
\end{subequations} 
for the translational $\rvec_i$ and orientational degrees of freedom $\phi_i$ of particle $i$, respectively. 
Forces $\textbf{f}_i$ and torques $M_i$ are derived from the colloid-colloid and coupling contributions to the free energy.
The Einstein relation applies for both the translational $D_{t}=k_BT /\gamma_t$ and  rotational diffusion constants $D_{r}=k_BT /\gamma_r$. 
The friction constants are $\gamma_t = 6\pi\eta_0 R$ and $\gamma_r = 8\pi\eta_0 R^3$ where $\eta_0$ is the viscosity of the medium.  
Similarly, both random noise parameters $\xi_t$ and $\xi_r$ satisfy the fluctuation-dissipation theorem.  
Each ABP self-propels with velocity $v_a$ with direction given by $\nhat_i=(\cos\phi_i,\sin\phi_i)$, which decorrelates in a rotational diffusion time scale $t_{rot}=D_{r}^{-1}$. 
It can be compared with the time scale in which an ABP moves its diameter $t_{swim}=2R /v_a$ to define the P\'{e}clet number $Pe=t_{rot}/t_{swim}$. 
Similarly, the persistence length $l_{pers}=v_a t_{rot}$ is the distance that the particle moves before decorrelating its orientation. 

Alignment interactions in active matter have been largely observed in bacteria systems capable of forming swarms. 
Polar aligning interactions have been widely modelled using the Vicsek model\cite{vicsek_novel_1995} (VM) which predicts the phase transition of randomly-oriented active particles into collective swarms.  
We introduce a Vicsek-like torque acting on particles $M_i^{cc}=K_{cc}/(\pi R_{cc}^2) \sum_j \sin(\phi_i-\phi_j)$ where particle \textit{i} interacts with neighbors $j$ within a cut-off distance which is set to\cite{sese-sansa_velocity_2018}  $R_{cc}=4R$. 
Similarly, a torque can be introduced to couple the orientation of ABPs with the BCP interface as $M_i^{cpl}=-K_{cpl}/(\pi R^2) \partial/\partial \phi_i \int d\textbf{r}  \left(\nabla\psi \cdot \mathbf{\hat{n}}_i \right)^2 $, to capture the alignment of ABPs interacting with walls\cite{palacios_guidance_2019}. 
We define the dimensionless torque parameters $g_{cc}=K_{cc}/(\pi R_{cc}^2 k_BT)$ and $g_{cpl}=K_{cpl}/(\pi R^2 k_BT)$ for the particle-particle and coupling contributions, respectively. 
The calibration of the effect of the aligning torques is shown in figure S1 \textbf{B} and \textbf{C} in the ESI.

On the other hand, the Cahn-Hilliard\cite{cahn_free_1959,cahn_free_1959-1} dynamics control the time evolution of the BCP order parameter 
$\partial\psi/\partial t=M\nabla^2 \left( \delta F_{tot} /\delta \psi \right)$ 
which drives the phase separation into ordered structures. 
The phenomenological mobility $M$ allows to define a BCP diffusive time scale $t_{BCP}\propto M^{-1}$ which in turn can be compared with the swimming time scale of the ABP to define a coupling P\'{e}clet-like dimensionless parameter $Pe^{cpl}=t_{BCP}/t_{swim}$. 
Additionally, the active energetic scale $\epsilon_{swim}=v_a\gamma_t (2R)$ can be compared with the colloid-colloid repulsive scale $\tilde{\epsilon}_{act}=\epsilon_{swim}/U_0$ and BCP-ABP coupling strength $\tilde{\epsilon}_{cpl}=\epsilon_{cpl}/\epsilon_{swim}$ where $\epsilon_{cpl}=\pi R^2 \sigma$. 
We select $R$ and $t_{rot}$ as the units of length and time, respectively.  
In this work, the colloid-colloid energetic scale $U_0$ is tuned to prevent overlapping between particles by maintaining $\tilde{\epsilon}_{act}\ll 1$. 
A complete description of the model used throughout this work can be found in the ESI. 

Several observables are used throughout this work to quantify the phase behaviour of the system. 
The modified nematic order parameter $S_{cpl}=\langle 2( \hat{\textbf{n}}_i \cdot \nabla \psi )^2-1 \rangle$ couples the ABP orientation with the vector pointing perpendicular to the interface $\nabla \psi$. 
The spatial inhomogeneity introduced by the BCP mesophase can prevent global polar ordering in the system.  
Therefore, the polarisation of the system is calculated as the average $P_{cc}=\langle P_{cc}^i \rangle$ of the local polar order for particle $i$ defined as $P_{cc}^i= 1/N_i \sum_{j=1}^{N_i} 
\hat{\textbf{n}}_i \cdot \hat{\textbf{n}}_j$ where the polarisation of particle $i$ is calculated with respect to the $N_i$ first neighbours within distance $r_{ij}^*=1.3 (2R)$. 
Similarly, the nematic-like particle-particle alignment is characterised as $S_{rel}=\langle S_{rel}^i \rangle$ with $S_{rel}^i= 1/{N_i} \sum_{j=1}^{N_i} 
2 \left (  
\hat{\textbf{r}}_{ij} \cdot \hat{\textbf{n}}_j
\right)^2-1$. 
The combination of $P_{cc}$ and $S_{rel}$ allows to distinguish between nematic and polar phases in 1D clusters. 
Additionally, cluster analysis allows to calculate the fraction $\Phi_{1D}$ of ABP clusters with 1D-like structure and the fraction $\Phi_{circle}^{BCP}$ of BCP domains with approximately circular shape.
A full list of the observables used in this work -as well as details on the calculation method- can be found in the ESI.




\begin{figure}[hbtp]
\centering
\includegraphics[width=1.0\linewidth]{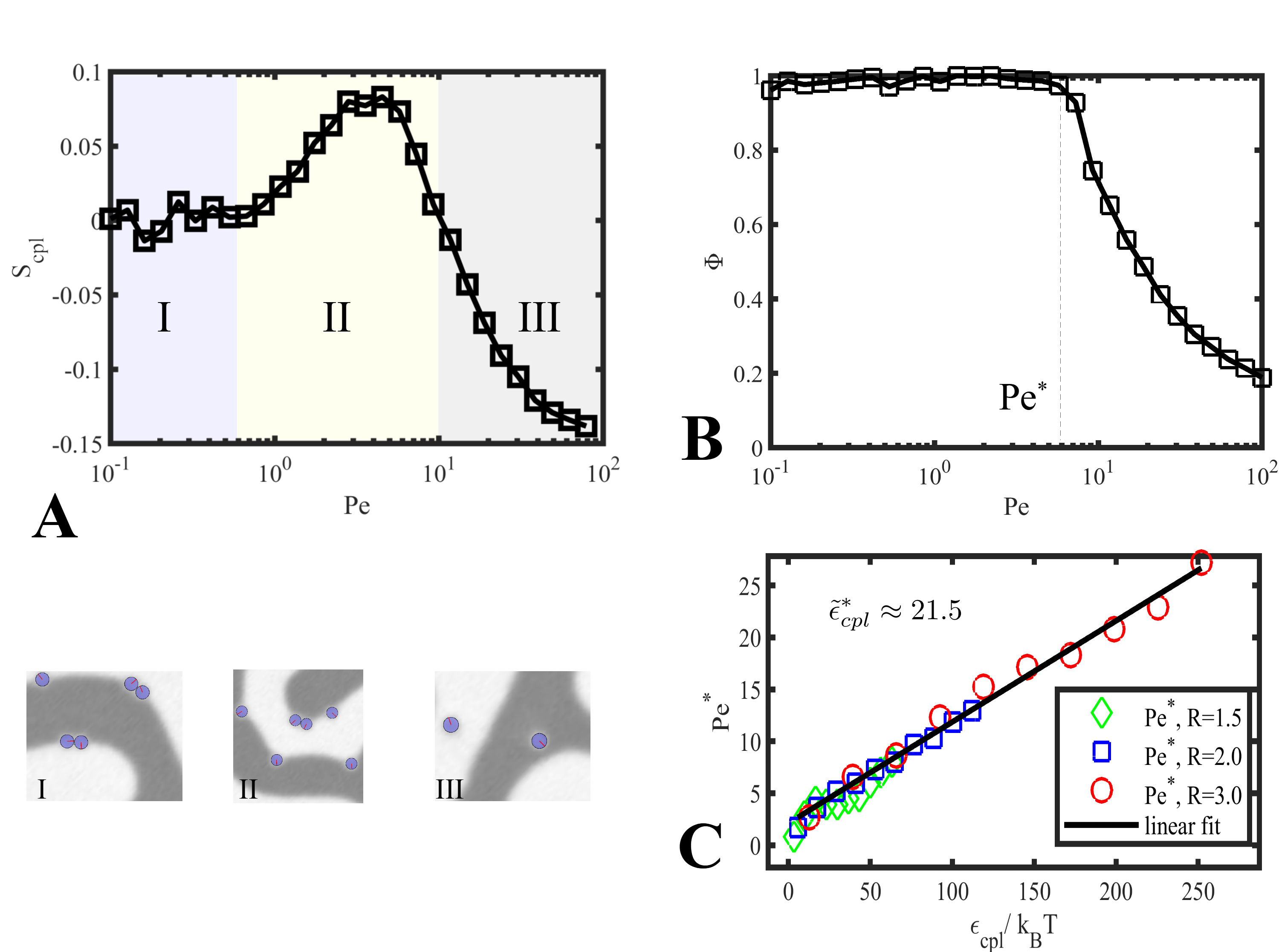}
\caption{
\textbf{Assembly of neutral active ABPs in a diluted regime ($\phi_p=0.01$)} in the absence of alignment torques $M_i^{cc}=M_i^{cpl}=0$. 
In \textbf{A} the nematic coupling order parameter $S_{cpl}$ is shown vs the P\'{e}clet number $Pe$ with particles sized $H_0/R=9.6$.
The sign of $S_{cpl}$ distinguishes $3$ regimes noted I, II and III with snapshots shown in the bottom-left of the panel.   
\textbf{B} shows the fraction $\Phi$ of colloids at the interface.
In \textbf{C} the critical P\'{e}clet collapses into a curve given by equation S22, for three particle radius. 
}
\label{fig:neutral.dilute}
\end{figure}

NPs  with a neutral interaction with both segments of  the BCP (surfactant-like, with $\psi_0=0$) anchor at the interface of the BCP domains\cite{kim_creating_2007}. 
This results in the soft confinement of neutral ABPs at the BCP interface. 
In figure \ref{fig:neutral.dilute} the ordering of ABPs is explored in terms of $Pe$, in a symmetric lamellar-forming BCP with $H_0/R=11.2$. 
No alignment torques are introduced regarding the particle-polymer coupling $g_{cpl}=0$ or the particle-particle polar alignmet $g_{cc}=0$.  
A low concentration $\phi_p=0.01$ can acquire both orientational and translational order, with three distinguishable regimes: 
\textbf{I} passive-like ABPs with low activity $Pe<1$ display no defined nematic order with respect to the interface $S_{cpl}\sim 0$ while anchoring at the interface $\Phi\sim1$. 
\textbf{II} moderately active ABPs with $1<Pe<10$ can self-propel along the interface before encountering a lamellar defect. 
ABPs that accumulate at lamellar defects contribute to positive nematic order, which couples the ABPs direction and the interface normal $S_{cpl}>0$, while the ABPs remain at the interface $\Phi\sim 1 $. 
\textbf{III} highly active partices with $Pe>10$ can detach from the interface when their swimming energy $\epsilon_{swim}$  is large enough to overcome the BCP coupling $\epsilon_{cpl}$.
In this regime, as the activity increases and more ABPs can escape the interface $\Phi<1$, the orientation of the ABPs is increasingly tangential to the interface $S_{cpl}<0$, as 
particles that approach an interface with a tangential orientation are likely to reside at the interface for a longer time than particles colliding \textit{head-on} with the interface.

The accumulation of ABPs at lamellar defects and their nematic ordering in regime \textbf{II} is a consequence of the fingertip-like morphology of symmetric BCP, that permits ABPs to accumulate at high-curvature regions of the interface. 
\footnote{This is a metaestable state, in which the lamella typically organise in experiments in the absence of external fields}
In the absence of defects in the lamellar morphology, \textit{i.e.} an equilibrated phase, the nematic order is not present as shown in figure S2, while BCP mesophases with intrinsic curvature display an increased peak. 
These results indicate the role of heterogeneity in complex medium leading to co-assembly of ABPs within BCP melts.

In figure \ref{fig:neutral.dilute} \textbf{B} a critical P\'{e}clet $Pe^*$ can be identified as the approximate value at which $\Phi\approx 0.9$.
Larger values of $Pe>Pe^*$ lead to an abrupt decrease of $\Phi$ as neutral particles are de-coupled from the interface. 
Exploring a wide range of $\sigma$ and $R$ values we identify the critical $Pe^*$  under different $\epsilon_{cpl}$ conditions. 
These data is seen to collapse into a single curve, following the dimensionless parameter $\tilde{\epsilon}_{cpl}^*$ dependence. 
This suggests that a critical dimensionless coupling value of the activity can be identified as $\tilde{\epsilon}_{cpl}^* \approx 21.5$, 
which is the critical value at which the self-propulsion is able to detach the particle from the interface trapping. 



\begin{figure*}[hbtp]
\centering
\includegraphics[width=0.99\linewidth]{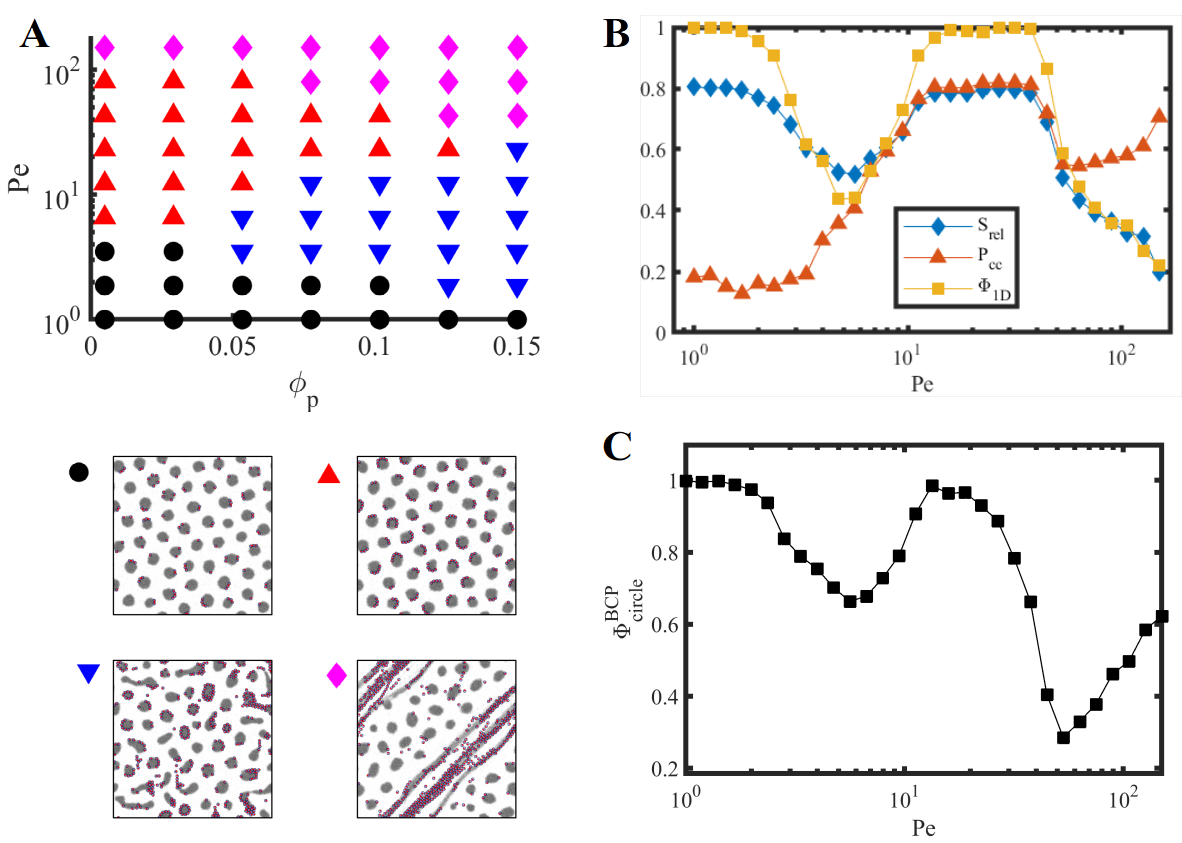}
\caption{
\textbf{Phase behaviour of neutral active particles with  concentration $\phi_p$}
with coupling alignment between colloids and BCP $g_{cpl}=34.9$ and polar alignment between particles $g_{cc}=0.06$ ,in \textbf{A}. 
Phases are marked as 
passive-like in black circles, 
downwards blue triangles for unstructured clusters, 
upwards red triangles for polarised trains at the interface, and 
diamonds for swarms escaped from the interface.  
In \textbf{B} and \textbf{C} the observable curves are shown for the ABPs and the BCP, respectively, for a fixed concentration $\phi_p=0.0875$.
Representative snapshots of each are shown in the bottom-left. 
}
\label{fig:neutral.phd}
\end{figure*}

At moderate ABP concentrations collective behaviour in active systems can be expected to emerge\cite{peterson_vesicle_2021}. 
Figure \ref{fig:neutral.phd} shows the  phase behaviour of finite concentrations $\phi_p$ of ABPs for an activity rate $Pe$, in the presence of strong particle-interface  $g_{cpl}=34.9$   and  moderate  particle-particle coupling $g_{cc}=0.06$ alignment. 
Neutral ABPs segregate towards the interface of circle-forming asymmetric BCP, with the system being initialised with ABPs at the interface of an already-equilibrated BCP.  
Several regimes can be distinguished in figure \ref{fig:neutral.phd} \textbf{A}, based on the structures formed by ABPs and the morphology of the BCP.
The curves of the BCP and ABPs observables are shown respectively in \textbf{B} and \textbf{C}, for a fixed concentration $\phi_p=0.0875$.

In figure \ref{fig:neutral.phd} \textbf{A} for low $Pe$ ABPs can flow within the interface before encountering an additional ABPs -marked as black circles- , leading to colloidal aggregation at the interface similar to the accumulation mechanism described in figure \ref{fig:neutral.dilute}. 
This low activity region is characterised by a 1D colloidal cluster shape $\Phi_{1D}\sim 1$ and
high nematic $S_{rel}\sim  0.8$ but low polar $P_{cc}\sim 0.2$ interparticle alignment, as can be seen in figure \ref{fig:neutral.phd} \textbf{B}, due to the low $g_{cc}$ coupling (see ESI video \path{neutral-phase-I_Pe1.mp4}). 
Low NP activity introduces no impact in the BCP morphology which remains circular, as shown in \textbf{C}. 

Higher activities $Pe$ allow ABPs to overcome the energetic barrier at the interface and escape into bulk regions -marked as blue and red triangular symbols . 
The required activity rate decreases as the ABP  concentration increases, indicating that ABPs escape collectively due to the particle-particle polar alignment. 
Furthermore, two regimes can be distinguished: 
unstructured clusters -marked as downwards blue triangles- are found for larger concentrations and moderate activity, where ABPs possess enough active energy to escape the interface but, upon entering the bulk regions, tend to form unstructured clusters characterised by a decrease in $P_{cc}$, $\Phi_{1D}$ and $S_{rel}$, as shown in figure \ref{fig:neutral.phd} \textbf{B}.
The relative high local concentration and the moderate activity promote the aggregation, which introduces considerable changes in the BCP mesophase, as shown in the decrease in circular BCP domains in figure \ref{fig:neutral.phd} \textbf{C} (see ESI video \path{neutral-phase-II_Pe8.mp4} ). 

\begin{figure}
    \centering
    \includegraphics[width=1.0\linewidth]{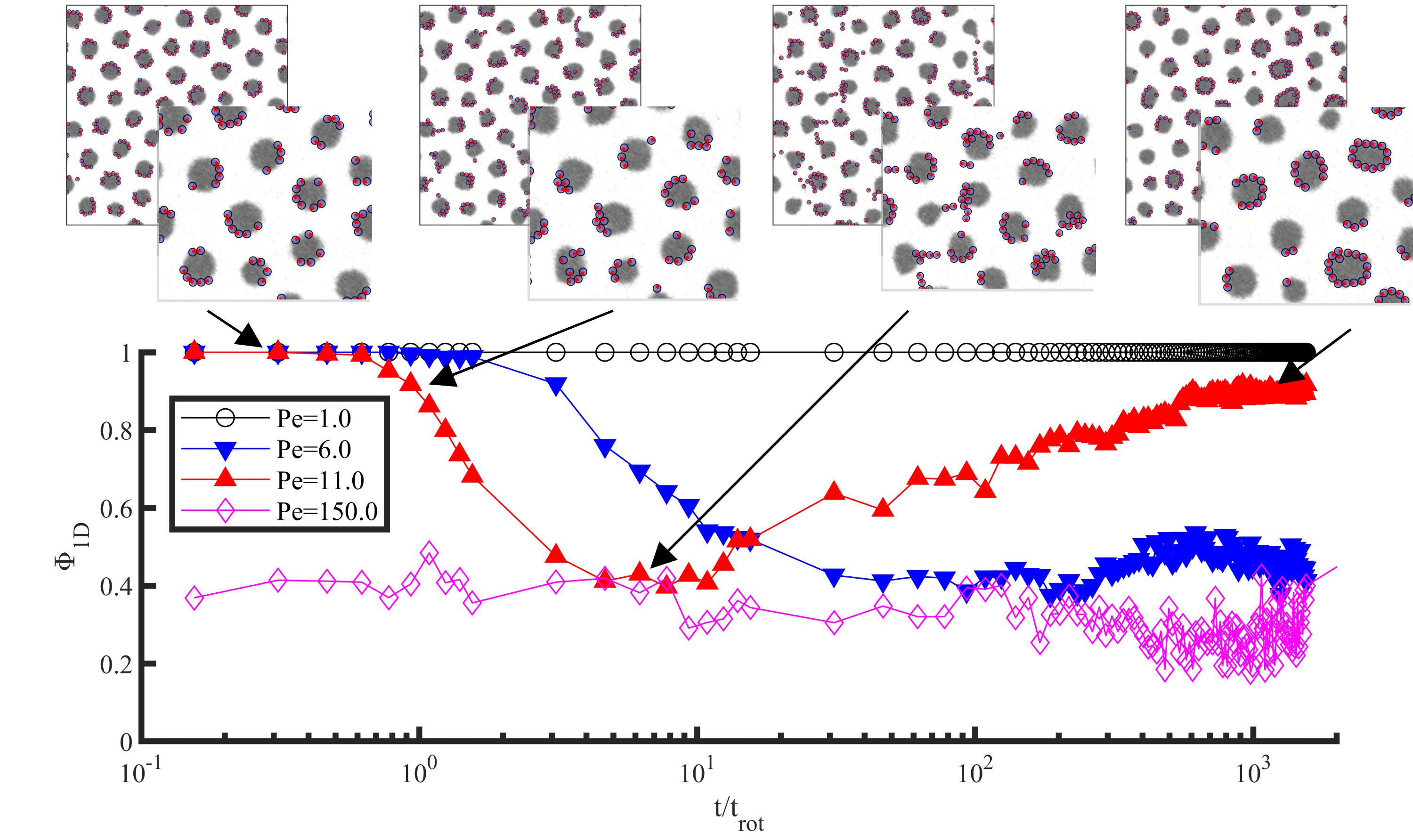}
    \caption{\textbf{Dynamic behaviour of the fraction of 1D ABP clusters} for four representative activities corresponding to the four regimes identified in figure \ref{fig:neutral.phd}.
    The system is initialised at $t=0$ with ABPs placed at the interface of equilibrated circle-forming BCP domains. 
    It can be compared with figure S3  for an initially-random condition. 
    }
    \label{fig:neutral.fraction.time}
\end{figure}

For moderate activities but lower concentrations ABPs can form polarised 1D clusters that maintain a continuous flow along the interface -marked as red upwards triangles. 
In this regime ABPs have enough active energy to not only escape the interface, but also to self-propel through the bulk phase before entering an additional BCP domain (see ESI video \path{neutral-phase-III.mp4} ).
This process can be repeated until the formation of polarised train-like clusters, characterised with high polarisation in \textbf{B}, in contrast to the low activity regime. 
The short-time dynamics of both train-like and unstructured cluster regime are considerably similar, as depicted in figure \ref{fig:neutral.fraction.time}, where a transient regime of lowering in the number 1D clusters is observed. 
Nonetheless, for higher activity rate ($Pe=11$, red upward triangles) ABPs are able to enter new BCP interfaces and form 1D polarised trains, while lower activity promotes the aggregation of ABPs into polarised clusters, leading to unstructured clusters ($Pe=6$, blue downwards triangles).

Finally, for considerably large activity , ABPs can be found completely detached from the interface, at the top of the phase diagram in figure \ref{fig:neutral.phd}. 
ABPs can re-orient upon collisions to form polarised clusters -blue diamonds- which leads to high polarisation but low interface-coupled nematic order, as shown in \textbf{B}. 
In this regime the activity is large enough to largely decouple the dynamics of the BCP and the ABPs, signalled by the relative time scales $Pe^{cpl}=0.13Pe$ and the energetic coupling $\tilde{\epsilon}_{cpl}=229/Pe$. 
As a result, the BCP relaxation time cannot accommodate the fast dynamics of the ABPs. 
In figure \ref{fig:neutral.phd} \textbf{C} the BCP mesophase is largely distorted as NPs are able to penetrate the interface, leading to a pronounced decrease in the fraction of circular domains for $Pe\sim 50$, which is followed by an increase in $\Phi_{circle}^{BCP}$ indicating the recovery of the equilibrium circular phase as the dynamics of the BCP and NPs are more decoupled.


Contrary to neutral NPs, selective NPs ($\psi_0=-1$) are miscible within one of the BCP phases\cite{kim_effect_2006}. 
When NPs are compatible with the minority phase in a circle-forming BCP, the BCP interface acts as a soft wall inducing confinement on the NPs. 
The strength of the confinement can be roughly estimated comparing the coupling energy and the swimming one, $\tilde{\epsilon}_{cpl}$. 
Particle-polymer and particle-colloid alignment are introduced, respectively with $g_{cpl}=34.9$ and $g_{cc}=0.06$ 
.

\begin{figure}
    \centering
    \includegraphics[width=0.99\linewidth]{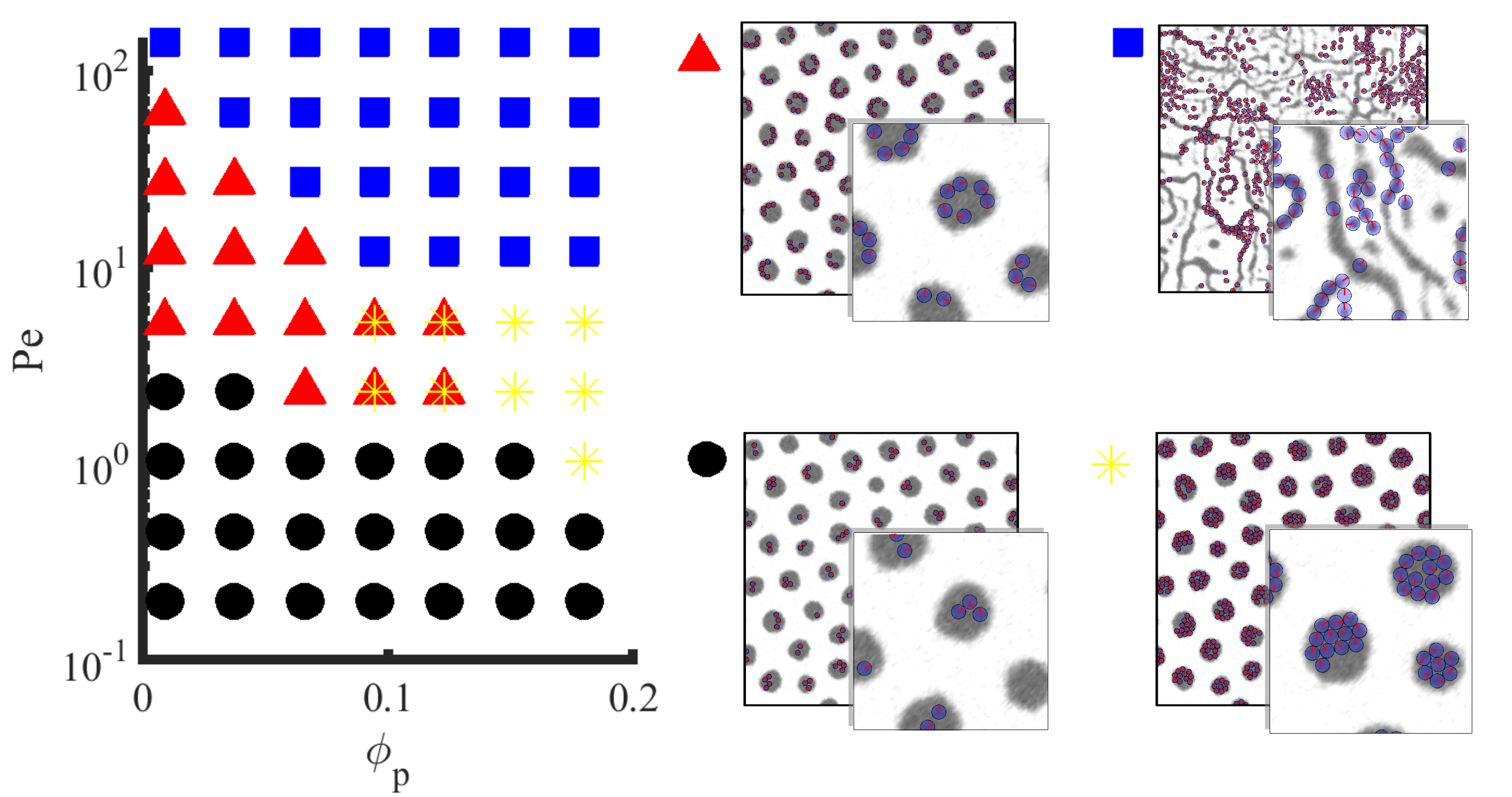}
    \caption{\textbf{Phase behaviour of selective active particles with a finite concentration $\phi_p$}. 
    Active particles align with respect to the BCP interface with $g_{cpl}=34.9$ and $g_{cc}=0.06$. 
    Markers indicate the colloidal regime: 
    disordered colloids within minority domains marked as black circles, 
    chain/ring structures marked with red triangles, 
    close-packed structures within minority domains marked as yellow asterisks and 
    free particles marked as blue squares. 
    }
    \label{fig:selective.phd}
\end{figure}

In figure \ref{fig:selective.phd} several co-assembled regimes can be identified, with passive-like ABPs marked as black circles in the low activity regime  
, displaying no  ordering. 
As the activity grows, the self propulsion leads to NPs accumulating at the domain walls, where they experience alignment torques promoting the tangential orientation of NPs with respect to the domain walls.
This corresponds to active persistence length $l_{pers}$ comparable to the diameter of the BCP domains (see figure S5 in the ESI).

For small concentrations, this leads to the additional re-alignment of NPs after colloid-colloid collisions, while the wall-particle coupling torque dominates. 
As a result, polarised 1D train-like colloids rotate tangentially to the domain walls, while softly confined within the circular phase, marked as red upside triangles. 
In this regime the domain wall coupling remain high enough to prevent NPs penetrating the interface, quantified by $\tilde{\epsilon}_{cpl}=229.4/Pe$, which assures that the BCP domains remain approximately circular. 
The formation of vortexes of active particles under confinement has been shown for active filaments in deformable vesicles\cite{peterson_vesicle_2021} or under geometric confinement\cite{woodhouse_spontaneous_2012,wioland_confinement_2013,lushi_fluid_2014,wu_transition_2017}.

For higher concentrations $\phi_p>0.09$, the crowded environment of NPs within some BCP domains limits the ability of NPs to rotate in train-like polarised clusters. 
Instead, 2D clusters can be formed, which may coexist with train-like clusters in other domains, depending on the  number of NPs within the local BCP domain. 
This regime is marked with yellow asterisks which can be seen to overlap with the train-like vortexes. 
In the onset of NP escaping, active colloids can considerably deform the BCP domain shape while ABPs can accumulate forming caps, in a similar way as described for active filaments\cite{peterson_vesicle_2021}.

For higher activity and larger concentrations, NPs can escape the soft confinement, marked as blue squares, and form swarm structures with internal polar ordering, similar to the ones shown in figure \ref{fig:neutral.phd}. 
Similarly, higher concentrations reduce the required activity $Pe$ to allow ABPs to escape the soft confinement. 
In this regime, the BCP dynamic is largely de-coupled with the ABPs one, as the swimming time scale is considerably smaller than the relaxation time of the BCP, following $Pe^{cpl}=0.13 Pe$.

\begin{figure}
    \centering
    \includegraphics[width=0.99\linewidth]{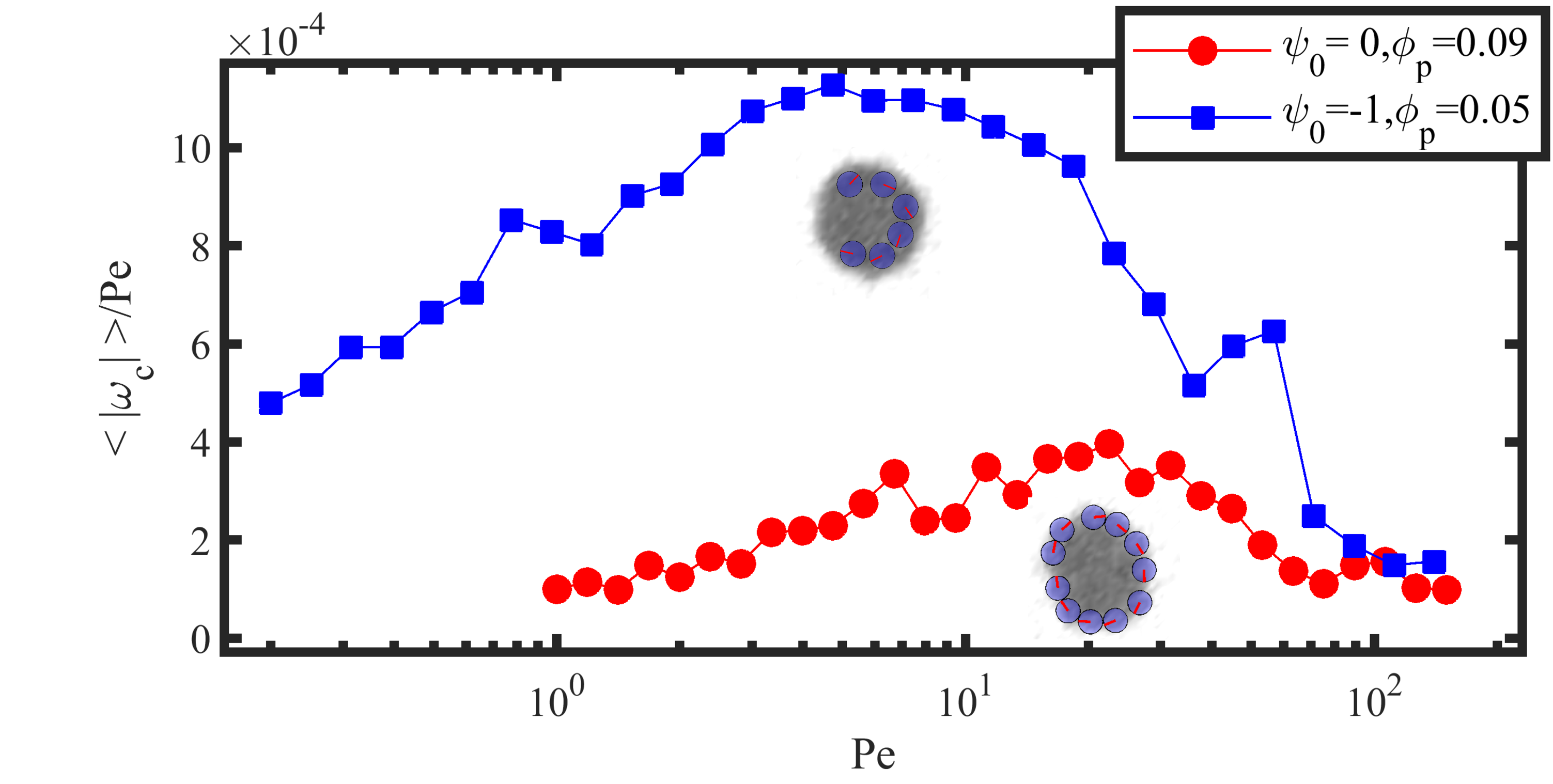}
    \caption{
    \textbf{Mean absolute value of the angular frequency }$\omega_c$ of ABPs with neutral ($\psi_0=0$, $\phi_p=0.08$) and selective interaction ($\psi_0=-1$, $\phi_p=0.05$) with the BCP mesophase. 
    }
    \label{fig:omega}
\end{figure}

Polarised train-like morphologies have been found for both neutral and selective ABPs, where the soft confinement of the curved BCP domains induce the emergence of rotational motion. 
Figure \ref{fig:omega} shows the appearance of chiral motion controlled by the BCP mesophase quantified by the angular frequency of ABPs.

ABPs have been shown to co-assemble within BCP melts to produce emergent organised structures displaying a rich phase behaviour.
This is a consequence of both the BCP intrinsic ordering and the self-propulsion of the ABPs subject to soft confinement by the BCP.  
Furthermore, high NP activity can lead to new co-assembled BCP morphologies resulting from the competition of the BCP and NP time and energetic scales. 
BCP melts are particularly well-suited to control the assembly of active particles as an example of nanostructured medium capable of inducing soft confinement.  


\begin{acknowledgments}
 The authors thank the European Union’s
Horizon 2020 Research and Innovation Programme project
VIMMP under grant agreement No 760907. 
I.P. acknowledges support from Ministerio de Ciencia, Innovación y
Universidades MCIU/AEI/FEDER for financial support under
grant agreement PGC2018-098373-B-100 AEI/FEDER-EU, from
Generalitat de Catalunya under project 2017SGR-884, and Swiss
National Science Foundation Project No. 200021-175719 .
\end{acknowledgments}

\bibliography{references}

\end{document}